\documentstyle[aps,prb,psfig]{revtex}

\begin{document}
\draft
\title{Microscopic study of electrical transport through individual 
molecules with metallic contacts: I. ``Band'' lineup, voltage drop and 
high-field transport}

\author{ Yongqiang Xue ~\cite{Xue} and Mark A. Ratner}
\address{Department of Chemistry and Materials Research Center, 
Northwestern University, Evanston, IL 60208}
\date{\today}
\maketitle

\begin{abstract}
We present the first in a series of microscopic studies of 
electrical transport through individual molecules with metallic contacts. 
We view the molecules 
as ``heterostructures'' composed of chemically well-defined atomic groups,  
and analyze the device characteristics in terms of the charge and potential 
response of these atomic-groups to the perturbation induced by the 
metal-molecule coupling and the applied electrical field, which are modeled 
using a first-principles based self-consistent matrix Green's function 
(SCMGF) method. As the first example, we examine the devices formed by 
attaching two benzene-based molecular radicals--phenyl dithiol (PDT) and 
biphenyl dithiol (BPD)--symmetrically onto two semi-infinite gold electrodes 
through the end sulfur atoms. We find that both molecules acquire a fractional 
number of electrons with similar magnitude and spatial distribution upon 
contact with the electrodes. The charge transfer creates 
a potential barrier at the metal-molecule interface that modifies 
significantly the frontier molecular states depending on the corresponding 
electron density distribution. For both molecules, the metal Fermi-level 
is found to lie closer to the highest-occupied-molecular-orbital (HOMO) 
than to the lowest-unoccupied-molecular-orbital (LUMO). Transmission 
in the HOMO-LUMO gap for both molecules is due to the metal-induced 
gap states arising from the hybridization of the metal surface states with the 
occupied molecular states. Applying a finite bias voltage leads to 
only minor net charge injection due to the symmetric device structure 
assumed in this work. But as current flows, the electrons within the 
molecular junction redistribute substantially, with resistivity 
dipoles developing in the vicinity of potential barriers. Only the 
delocalized $\pi$-electrons in the benzene ring can 
effectively screen the applied electric field. For the PDT molecule, 
the majority of the bias voltage drops at the metal-molecule 
interface. But for the BPD molecule, a significant amount of the 
voltage also drops in the molecule core. The field-induced 
modification of the molecular states (the static Stark effect) becomes 
significant as the bias voltage increases beyond the linear-transport 
region. A bias-induced reduction of the HOMO-LUMO gap is observed for 
both molecules at large bias. The Stark effect is found to be stronger for 
the BPD molecule than the PDT molecule despite the longer length of the 
former. For both molecules, the peaks in the conductance are due to 
electron transmission through the occupied rather than the unoccupied 
molecular states. The calculation is done 
at room temperature, and we find the temperature dependence of the 
current-voltage characteristics of both molecules is negligible.  
\end{abstract}

\date{\today}

\section{Introduction\label{S1}}
Exploring the use of individual molecules as active components in 
electronic devices~\cite{Ratner74} has been at the forefront of 
nanoelectronics research in recent years due to the potential advantage 
of ultrahigh density/speed 
and low-cost device fabrication through self-assembly/self-organization 
processes that such molecular-scale devices may bring~\cite{Ratner98,JGA00}. 
Numerous useful device characteristics including molecular 
rectifying diodes, negative differential resistance and field-effect 
transistors have been demonstrated using molecular-scale structures including 
small conjugated molecules~\cite{Reed97,Tour01}, 
single- and multi-wall carbon nanotubes~\cite{DekkerR99,NTFET} 
and macromolecules like DNA~\cite{Dekker00}. A quantitative 
understanding of the physical 
mechanisms underlying the operation of such diverse molecular-scale 
devices has been and remains a major challenge in nanoelectronics research. 

The electrical characteristics of molecular devices are usually measured by 
sandwiching the molecules between two metallic 
electrodes~\cite{JGA00,Reed97,Metzger97,Datta97,Xue991,Hong00,Bao02}. The measured 
transport characteristics reflects both the nature of the metal-molecule 
interface and the properties of the molecular 
layer~\cite{Xue01,Xue021,Datta01,Lang002,Lang01,Guo011,Brandbyge02,Nitzan,EK99,Hush00,Semi}. 
There are basically three electronic processes of interest in such 
metal-molecule-metal junctions: charge transfer between the metals and 
the molecules, the change of the electrostatic potential and the 
modification of the molecular geometry and electronic states. 
The nature of these processes under both zero and nonzero bias determines the 
electrical characteristics of the molecular junction. Here it is natural to 
separate the problem into device at equilibrium and device out of 
equilibrium since: (1) the linear dc-transport property of the molecular 
device probes the equilibrium charge distribution of the molecular 
junction established by adsorption onto the electrodes; (2) the nonlinear 
transport probes the charge response of the molecular junction to the 
applied electrical field established through a finite bias voltage. 
A microscopic theory of molecular electronics will therefore need to: 
(1) Determine the appropriate geometry of the molecular junction; 
(2) determine the self-consistent charge transfer and the resulting 
lineup of the molecular levels relative to the metal 
Fermi-level and the modification of the molecular states upon formation 
of the metal-molecule-metal junction; (3) determine the molecular 
screening of the applied electric field, the field-induced modification 
of molecular states (the static Stark effect) and the 
non-equilibrium electron distribution 
when current is flowing; (4) determine the current/conductance-voltage 
characteristics and their correlation with the molecular and device 
structures. None of these issues has been satisfactorily solved 
such that theory can be used in conjunction with experiment for an 
unambiguous identification of the device operation mechanism. The 
purpose of our work is to elucidate the above device electronic 
processes and their dependence on the given molecular and device 
structures under both equilibrium and nonequilibrium conditions 
within a well-defined theoretical model. Specifically, we will 
investigate the different 
aspects of molecular transport due to the interplay between the 
electronic processes at the metal-molecule interface, the 
electronic processses in the molecule core and the molecular response to 
the applied electrical field when current is flowing. Correspondingly, we 
focus our attention on the electron dynamics and calculate the device 
characteristics within the coherent transport regime. Our emphasis is on 
the conceptual understanding and the chemical trends obtained from detailed 
microscopic calculation of simple but representaive molecular and device 
structures. Our work can therefore be considered as a ``minimal'' 
quantitative microscopic model of single-molecule electronics. 

The ionic dynamics of the metal-molecule-metal junction 
can also be affected by the above electronic processes both at 
equilibrium and out of equilibrium, i.e., the adsorption and 
bias/current-induced conformation change. Neither aspect will be 
treated here. By confining ourselves to the coherent transport regime, 
we also neglect the effect of electron-vibrational coupling on 
transport~\cite{Ho98}. Solving the adsorption-induced 
conformation change requires an accurate knowledge of the surface lattice 
structure at the atomic-scale for devices under applied voltage. This is 
almost never known in typical molecular transport measurement. Since 
one goal of such calculation is to 
provide the input nuclear configuration for transport calculation, our 
purpose can be served equally well by examining the effect that different 
adsorption and molecular geometries may have on the device characteristics, 
which will be treated in subsequent papers of the series. A rigorous 
solution of the bias/current-induced conformational change and the general 
solution of inelastic contribution to non-linear current due to 
electron-vibrational coupling is a challenging topics 
which requires solving the nonequilibrium dynamics of the coupled 
electron-nuclei systems~\cite{Yu99,Fisher01} since the molecular potential 
energy surface is affected by the nonequilibrium electron 
dynamics and electron energy is dissipated in moving the 
atoms. By focusing on the electron dynamics in the molecular junction with 
fixed geometry, the present work will provide a reference for 
evaluating the importance of such additional complications and provide the 
necessary input for further investigation in situations where they are 
indeed critical. 

In typical molecular junction measurement, conjugated molecules are 
attached to the metallic electrodes through appropriate 
end groups. In this paper we will consider current 
transport through two benene-based molecular radicals--phenylene dithiol (PDT) 
and biphenylene dithiol (BPD)--adsorbed symmetrically onto two 
semi-inifinite gold $<$111$>$ electrodes through the end sulfur atoms. 
These structures chosen are among the simplest possible but are 
still representative of current experimental work. 
For the device at equilibrium, we obtain the self-consistent charge 
transfer, the adsorption-induced change in the electrostatic potential, 
the lineup of the molecular level relative to the metal Fermi-level and 
the modification of the individual molecular states due to the 
metal-molecule coupling. For the device out of equilibrium, we obtain the 
charge injection/redistribution within the metal-molecule-metal junction, 
the molecular screening of the applied field and 
the resulting voltage drop, the modification of the molecular states by 
the applied field, the non-equilibrium occupation of molecular orbitals, the 
current/conductance-voltage characteristics for the given contact geometry. 

The calculation we present is performed using a recently developed 
self-consistent matrix Green's function (SCMGF) method~\cite{Xue01,Xue021} 
which is based on the non-equilibrium generalization of the 
quasi-particle Green's function theory~\cite{Keldysh65,Mahan87,HJBook} 
and uses a finite local-orbital basis set. By replacing the quasi-particle 
exchange-correlation self-energy~\cite{Wilk99} with the exchange-correlation 
potential within the density functional theory~\cite{LMBook,DGBook}, the 
well-established technique of self-consistent field theory of 
molecular electronic structure can then be utilized for transport 
calculations~\cite{Xue021}. The real-space formulation 
of our approach allows us to provide an intuitive and coherent physical 
picture of the molecular tranport by analyzing the device elctronic 
processes both at the atomic-scale and on the basis of individual 
molecular orbitals. We view the molecules 
as comprised of chemically well-defined atomic groups and  
interpret their electrical characteristics in terms of the response of 
these atomic groups to the perturbation induced by the metal-molecule 
coupling and the applied electric field. We emphasize the insight 
obtained with such atomic-scale analysis 
and show the important effect of atomic-scale charge and potential 
inhomogeneity on device characteristics at the molecular scale, which may 
otherwise be obscured by the molecular-level analysis treating the 
molecules as a whole. In particular, two important conclusions come out of 
this work: (1) the adsorption-induced modification of molecular states is 
larger than the field-induced effect unless we go to large bias, so  
accurate modeling of the electronic processes at equilibrium 
is critical for determining the low-bias transport characteristics; 
(2) the effect of both the metal-molecule coupling and the applied 
electric field in turning the individual molecular orbitals into effective 
conduction channels depends on the detailed charge and potential 
distribution across the metal-molecule-metal junction and may be different 
for different molecular orbitals. ``Engineering'' the charge and potential 
inhomogeneity as commonly done in quantum semiconductor heterostructures 
will be equally important at the molecular scale. Both aspects 
have been largely neglected in the past. We therefore hope the results 
obtained here will provide a useful guide regarding the nature of 
electron transport at the ultimate limit of device scaling and the 
prospect of device engineering through molecular design. 

\section{Theoretical Model \label{S2}}
\subsection{The self-consistent matrix Green's function theory}
The details of the self-consistent matrix Green's function theory have  
been described elsewhere~\cite{Xue01,Xue021}, so we only give a brief 
summary here to clarify the assumptions and approximations and to show 
how the physical observables are computed.   
Similar but different methods based on the NEGF approach have been developed 
independently by other research groups~\cite{Datta01,Guo011,Brandbyge02}. 
We feel that this approach provides a natural link between quantum 
transport, first-principles electronic structure theory and qualitative 
molecular orbital theory~\cite{Hoffmann88,ABW85}. 
 
Due to metallic screening in the electrodes, the charge and potential 
perturbations induced by molecular adsorption extend only over a 
finite region into the metal surface~\cite{Lang82,BT99}. We define 
an ``extended molecule'' which includes both the molecule and the surface 
atoms perturbed by molecular adsorption~\cite{Ratner99,Joachim97}. The 
surface atoms also form an adiabatic reflectionless contact with the rest 
of the electrodes, which can then modeled as infinite electron reservoirs 
commonly assumed for mesoscopic transport problems~\cite{Xue021,Landauer}. 
The size of the ``extended molecule'' is chosen such 
that charge neutrality is maintained approximately under both zero and finite 
biases. The central quantities in the NEGF theory are the retarded and 
correlation Green's function $G^{r}$ and $G^{<}$~\cite{HJBook,Mahan87}. 
We expand both the wavefunctions $\psi_{\mu}$ and the Green's functions 
using a finite set of local orbital functions $\phi_{i}$, 
\begin{eqnarray}
\label{Gexp}
G^{r;\sigma}(\vec r,\vec r';E)
&\cong& \sum_{i,j}G_{ij}^{r;\sigma}(E) 
  \phi_{i}(\vec r) \phi_{j}^{*}(\vec r'), \\
G^{<^;\sigma}(\vec r,\vec r';E)
&\cong& \sum_{i,j}G_{ij}^{<;\sigma}(E) \phi_{i}(\vec r) \phi_{j}^{*}(\vec r')
\end{eqnarray} 
where $\sigma$ is the spin index. 
We obtain the retarded and correlation matrix Green's function 
by solving the Keldysh-Kadanoff-Baym (KKB) equation in the matrix 
form~\cite{Xue021}: 
\begin{eqnarray}
G_{MM}^{r;\sigma} &=& \{ E^{+}S_{MM}-H_{MM}^{\sigma}-V^{ext;\sigma}_{MM}
-\Sigma_{L}^{r;\sigma}(E)-\Sigma_{r}^{r;\sigma}(E)   \}^{-1}, \\
G^{<;\sigma}(E)
    & =& i[G^{r;\sigma}(E)\Gamma_{L;\sigma}(E)G^{a;\sigma}(E)]f(E-\mu_{L})
    +i[G^{r;\sigma}(E)\Gamma_{R;\sigma}(E)G^{a;\sigma}(E)]f(E-\mu_{R})
\label{KKB}
\end{eqnarray}      
where the effect of the contact is incorporated as the self-energy operators 
$ \Sigma_{L(R);\sigma}^{r}$,
\begin{eqnarray}
\Sigma_{L}^{r;\sigma}(E) 
&=& (E^{+}S_{ML}-H_{ML;\sigma}) 
  G^{0r;\sigma}_{LL} (E^{+}S_{LM}-H_{LM}^{\sigma}), 
\nonumber \\
\Sigma_{R}^{r;\sigma}(E) &=& (E^{+}S_{MR}-H_{MR}^{\sigma}) 
G^{0r;\sigma}_{RR} (E^{+}S_{RM}-H_{RM}^{\sigma}),
\nonumber \\
\Gamma_{L(R)}^{\sigma} &= & i(\Sigma^{r;\sigma}_{L(R)}
                      -(\Sigma^{r;\sigma})^{\dagger}_{L(R)}).
\end{eqnarray}
Here $G^{0r;\sigma}_{LL(RR)} = 
(E^{+}S_{LL(RR)}-H_{LL(RR)}^{\sigma})^{-1}$ are  
the surface Green's functions of the left (L) and right (R) contacts 
(the electrodes with the perturbed surface atoms {\emph removed}) 
which can be obtained from that 
of the semi-infinite surface. $H_{MM}^{\sigma}$ represents 
the part of the Fock matrix contributed by the charge distributions 
(both nuclei and electron) in the ``extended molecule'' only, while 
$V^{ext;\sigma}(\vec r)$ represents the long-range coulomb potential due to 
the equilibrium charge (ionic and electronic) distribution in the contact 
region, which includes the linear voltage drop due to the applied bias. 
The $S$ are overlap matrices. 
The applied potential $V^{ext}$ changes the charge and potential in the 
``extended molecule'', requiring a self-consistent solution. 

Given the electron density or the density matrix in the 
``extended molecule'', the calculation of the Fock matrix 
$H_{MM;\sigma}$ is the same as that of standard molecular electronic 
structure calculation, which greatly facilitates the implementation of our 
procedure using standard molecular electronic structure codes~\cite{Xue021}.  
The self-consistent calculation proceeds by computing the input density 
matrix to the next iteration from the correlation matrix Green's function 
computed in the current iteration:
\begin{equation}
\label{DMNE}
\rho_{ij}^{\sigma}=\int \frac{dE}{2\pi i}G^{<;\sigma}_{ij}(E)
\end{equation}
\emph{The density matrix is simply the energy integration of the 
matrix correlation function}. The integration over energy can be performed 
conveniently in the complex energy plane~\cite{Xue021,Dede82}. Once the 
self-consistent calculation converges, we can calculate all the physical 
observables from the matrix retarded Green's function and 
the density matrix. 

\subsection{Analyzing molecular transport}
We calculate the charge density using 
\begin{equation}
\label{Den}
\rho(\vec r)=\sum_{ij;\sigma} \rho_{ij}^{\sigma} 
\phi_{i}(\vec r) \phi_{j}^{*}(\vec r),    
\end{equation}
from which we can obtain the electrostatic potential in the molecular 
junction from the Poisson equation. We can also calculate 
charges associated with each atom using Becke's atomic-partition 
scheme~\cite{Becke88CP},
\begin{equation}
N=\int d\vec r \rho(\vec r) 
=\sum_{i} \int d\vec r W_{i}(\vec r)\rho(\vec r)
=\sum_{i}N_{i}
\end{equation}
Here the atomic weight function $W_{i}(\vec r)$, which satisfies 
$\sum_{i}W_{i}(\vec r)=1$ everywhere in space, is determined 
such that it is centered on the atom $i$ and is non-negligible only in 
a region close to its atomic center~\cite{Becke88CP}. 

The local density of states (LDOS) gives the energy-resolved charge 
density distribution:
\begin{equation}
\label{LDOS}
n^{\sigma}(\vec r,E)=-\frac{1}{\pi} \lim_{\delta \to 0^{+}} \sum_{ij}
Imag[G_{ij}^{R;\sigma}(E+i\delta)] \phi_{i}(\vec r) \phi_{j}^{*}(\vec r), 
\end{equation}
The spatial integration of LDOS gives the density of states,
\begin{equation}
\label{DOS}
n^{\sigma}(E)=\int d\vec r n^{\sigma}(\vec r,E)
= -\frac{1}{\pi} \lim_{\delta \to 0^{+}} 
  Tr\{Imag[G^{R;\sigma}(E+i\delta)] S\}, 
\end{equation}
To identify the contribution of 
the individual molecular orbitals to the total density of states, 
we project it onto the basis of molecular orbtias. This is done by 
transforming $G^{R}$ into the basis of the molecular orbitals obtained 
by diagonalizing the (self-consistent) Fock matrix corresponding to the 
molecule. The imaginary part of the diagonal element of the transformed 
$G^{R}$ gives the projected density of states (PDOS) of the corresponding 
molecular orbitals. The peak position of the PDOS characterizes the perturbed 
molecular energy level, while the broadness reflects the coupling strength 
of the molecular oribitals with the metallic surface states.

A critical question in molecular transport is how the electron 
occupation of the molecular orbitals changes as the molecule is driven 
out of equilibrium by a finite bias voltages. The information is contained 
in the non-equilibrium density matrix $\rho_{ij}^{\sigma}$ 
(Eq.\ \ref{DMNE}). Transforming to the basis of molecular orbitals, the 
diagonal elements of the density matrix give the the non-equilibrium 
electron occupation of the corresponding molecular levels.  

A general formula for the current through a mesocopic system with arbitrary 
interaction in contact with two non-interacting electrodes is: 
\begin{eqnarray} 
\label{MWCurrent}
I &=& \frac{e}{h}\int dE \sum_{\sigma} 
Tr \{ [(\Gamma_{L}^{\sigma}-\Gamma_{R}^{\sigma})(E,V)\ 
iG^{<;\sigma}(E,V)] \nonumber \\
  &+& [f(E-\mu_{L})\Gamma_{L}^{\sigma}(E,V)-f(E-\mu_{R}) 
\Gamma_{R}^{\sigma}(E,V)] A^{\sigma}(E,V)] \}
\end{eqnarray}
where $A^{\sigma}=i(G^{r;\sigma}-G^{a;\sigma})$ is the spectral function. 
Since the only scattering mechanism in the coherent transport 
regime is that by the contacts (introducing additional scattering mechanisms 
within the molecule will lead to additional terms in self-energy and current), 
we arrive at the familiar Landauer-type current formula: 
\begin{equation}
I=\frac{e}{h} \int dE \sum_{\sigma}T^{\sigma}(E,V)[f(E-\mu_{L})-f(E-\mu_{R})]
\label{IV}
\end{equation}    
where the transmission probability $T^{\sigma}$ through the 
molecule~\cite{MW92,Datta95,Xue021} is obtained from: 
\begin{equation}
\label{TEV}
T^{\sigma}(E,V)=Tr[\hat t^{\sigma}(E,V)]=
Tr[\Gamma_{L}^{\sigma}(E,V)G^{R;\sigma}(E,V)\Gamma_{R}^{\sigma}(E,V)
[G^{R;\sigma}]^{\dagger}(E,V)],
\end{equation}
Note that equations (\ref{IV}-\ref{TEV}) 
hold \emph{for both orthogonal and non-orthogonal 
basis functions}. To identify the contribution of 
individual molecular orbitals to the transmission, we transform the 
transmission matrix $\hat t^{\sigma}(E,V)$ into the the basis of 
the molecular orbitals whose diagonal elements give the transmission 
probability through the corresponding molecular orbitals. 

Combined with the spatially resolved LDOS and charge density, 
the total DOS, the transmission coefficient and their projection 
onto individual molecular orbitals provide a set of useful qualitative 
analysis tools to establish the connection between the molecular 
electronic structure and the transport characteristics of the 
metal-molecule-metal junction, similar to the use of qualitative molecular 
orbital theory in quantum chemistry~\cite{ABW85}. 
A caveat of projecting the physical observable onto the 
individual molecular orbitals is that phase interference information 
is lost. Only the summation over all molecular orbitals (conserved by 
the matrix transformation), not the individual components,  
corresponds to the quantum mechanical physical observable. 
The situation is similar to what we met when we try to decompose the 
electron density of the molecule into the contribution of individual 
atoms through a Mulliken-type population analysis or other charge 
partition schemes as we used here. Only the total density but not the 
atom-partitioned density corresponds to a physical observable.  

Within the coherent transport model, the only temperature 
effect is from the two electrode Fermi-Dirac distributions. We can 
separate the current into two components, the ``tunneling'' component 
$I_{tun}$ and the ``thermionic emission'' component $I_{th}$ as follows,
\begin{equation}
I = I_{tun}+I_{th} \nonumber 
  = \frac{e}{h} [\int_{\mu_{L}}^{\mu_{R}}+
  (\int_{-\infty }^{\mu_{L}}+\int^{+\infty }_{\mu_{R}})] 
 dE T(E,V)[f(E-\mu_{L})-f(E-\mu_{R})] 
\end{equation}

\subsection{Device model}

We choose the electrostatic potential in the middle of the (empty) 
bimetallic junction as the energy reference, then the equilibrium 
Fermi-level $E_{f}$ is fixed at the negative of the metal work 
function $-5.31 (eV)$ 
for single-crystal gold electrodes. The electrochemical 
potential of the two electrodes is fixed by the applied bias voltage 
$V$ at $\mu_{L}=E_{f}-eV/2$ and $\mu_{R}=E_{f}+eV/2$. Note 
\emph{the Fermi-level positions are fixed by the bimetallic junction 
alone without the molecular insertion}. The voltage drop across the 
molecular junction is determined by the 
molecular response to the metal-molecule coupling and the applied bias 
voltages. The bias polarity is chosen such that for positive bias 
the electron is injected from the right electrode. 
 
In this work, the electronic structure of the ``extended molecule'' 
is described using the Becke-Perdew-Wang(BPW91)~\cite{Becke88GGA,PW91} 
parameterization of the spin-polarized density-functional theory 
(SDFT)~\cite{LMBook,JG89} within the Generalized-Gradient 
Approximation (GGA)~\cite{PW91}. We also replace the atomic core by 
an appropriate pseudopotential~\cite{KBT} with the 
corresponding optimized Gaussian basis sets~\cite{Stevens84,Note1}. The 
geometry of the adsorbed molecule is taken to be the same as the 
singlet geometry of the free molecule optimized at the 
BPW91/$6-31G^{*}$ level (we asuume the bare molecule to be the molecular 
biradical with the end H atoms removed). The adsorption geometry is 
chosen such that the end atoms sit in front of the center of the triangular 
pad of the three gold atoms on the Au$<$111$>$ surface (the end 
sulfur-suface distance is $1.9 \AA$). Six nearest-neighbor gold 
atoms on each metal surface are included into the ``extended molecule''. 
Within the range of second-nearest-neighbor coupling, there are 
$12$ metal atoms in the 
first surface layer and $14$ metal atoms on the second surface layers 
on each side coupled to the ``extended molecule''. Only the blocks of 
the surface Green's function corresponding to these atoms are needed, which 
in turn can be calculated from the surface Green's function of the 
semi-infinite metal using tight-binding method parametrized by fitting 
accurate bulk band structure calculations~\cite{Xue01,Papa86}. The results 
and conclusions given in the following are not affected significantly by 
small changes in the molecular and adsorption geometry. 
The structures of the molecule junction are shown 
in Figs.\ (\ref{xueFig2-1-1}) and (\ref{xueFig2-1-2}). 
The calculation is performed using a modified version 
of GAUSSIAN98~\cite{Xue01,G98}.

\section{Device at equilibrium \label{S3}}

The problem at equilibrium is a generalization of the familiar 
chemisorption problem in surface science since two metallic surfaces are 
involved. For a molecular orbital to be turned into an effective 
conduction channel of the metal-molecule-metal 
junction (within the coherent transport regime), it needs to couple well 
to both electrode states. Its energy must lie within the energy window 
determined by the bias voltage and the thermal broadening through the 
electrodes Fermi-distribution. Both aspects are affected by the 
molecular adsorption, which may well be more significant 
than the subsequent application of the bias voltage, as detailed in 
subsequent sections.  
     
\subsection{The electronic structure of the molecules and the 
identification of chemical groups\label{S3-1}}

The transport property of the molecular junction is 
determined mainly by the hybridization of the surface metal states with 
the frontier molecular orbitals, i.e., the molecular states whose energies 
lie closest to the metal Fermi-level. So we start by illustrating  
in Fig.\ (\ref{xueFig3-1-1}) and Fig.\ (\ref{xueFig3-1-2}) the charge 
distribution of the HOMO-1,HOMO,LUMO and LUMO+1 states of the two isolated 
molecules (with the end H removed). For both molecules, the HOMO-1 
states are localized on the end sulfur atoms, the HOMO states are 
delocalized over entire conjugated backbone while the LUMO and the LUMO+1 
states are benzene based. The LUMO state of BPD also has a finite weight 
on the end sulfur atoms. Both molecules can be viewed as ``heterostructures" 
composed of the end sulfur atoms and the benzene rings. Due to the 
non-coplanar geometry of the two benzenes in BPD molecule (the torsion 
angle is $37^{\circ}$), the orbital overlap between the corresponding 
$\pi$ electrons is weak. Our analysis of the electrical transport through 
the two molecules will be based on the electrical response of these 
chemical units. 

\subsection{Charge transfer and electrostatic potential change 
in the molecular junction \label{S3-2}}

The adsoption-induced charge transfer across the metal-molecule interface   
is the central quantity for the device at equilibrium since the 
linear-response 
transport probes the equilibrium charge distribution of the molecular 
junction. The perturbation introduced by the metal-molecule coupling is 
largely a localized interaction. The corresponding charge-transfer process 
is determined by the interfacial chemistry and can be understood 
qualitatively from the local bonding analysis across the metal-molecule 
interface as in the chemisorption problem~\cite{Xue01,Hoffmann88}. 

Due to the identical end group and bonding configuration across the 
metal-molecule interface, both the magnitude and the spatial distribution 
of the transferred charge for the two molecules are quite similar. 
This is illustrated in 
Figs. (\ref{xueFig3-2-1}-\ref{xueFig3-2-2}), where we plotted 
the difference between the self-consistent charge distribution in the 
gold-molecule-gold junction and the charge distribution in the isolated 
molecule \emph{plus} the charge distribution of the isolated bimetallic 
contact. For clarity and to aid visualization, we have plotted both the 
regions where charge accumulation and depletion occur and the spatial 
distribution of the transferred charge as a function of position in the X-Y 
plane (defined by the left benzene ring). The charge transfer process 
involves mainly the end sulfur atoms 
and neighbor carbon atoms and decays rapidly as we move away from the 
metal-molecule interface. This is clearly seen for the BPD molecule where 
the charge perturbation induced by the two metal-molecule contacts decays 
into the interior of the molecule without effectively interfering with 
each other (Fig.\ \ref{xueFig3-2-2}). Note this is not necessarily 
due to the non-coplanar geometry. The decay of charge perturbation is 
already obvious for the PDT while for BPD it becomes negligible 
before reaching the inter-benzene bonding region. The number of 
electrons increases in the 
sulfur-gold bonding region due to the rehybridization of the sulfur $P_{z}$ 
orbital in forming gold-sulfur bond and the transfer of charge from 
gold atom to sulfur atom due to electronegativity difference. 
Electron density also increases in the neighbor carbon $P_{z}$ 
orbital since the HOMO level has large weights on both the sulfur and 
carbon $P_{z}$ orbitals. The electron density decreases in the 
sulfur-carbon $\sigma$-bonding region, the sulfur $P_{x}$ orbitals 
and also the gold $s$ orbitals. This is because the formation of surface 
gold-sulfur bond 
involves charges originally residing in the sulfur $P_{x}$ and gold $s$ 
orbitals thus weakening the bonding between sulfur and carbon. The 
direction of charge transfer is from the gold electrodes to the molecules. 

Accompanying the charge transfer across the gold-molecule interface, the 
electrostatic potential also changes. 
The difference of the electrostatic potential in the junction and the 
electronic potential in the isolated molecule \emph{plus} that in the 
isolated bimetallic contact gives the potential perturbation due to 
the formation of the contact. This is plotted in 
Figs. (\ref{xueFig3-2-3-1}) and (\ref{xueFig3-2-3-2}). 
The net transfer of electrons into the molecule increases the electrostatic 
potential inside the molecule creating a potential barrier between 
the metal surface and the end sulfur atom. There is also a potential well between the 
sulfur atom and the benzene ring because of the decrease of electron density 
in the sulfur-carbon bonding region. Although the pattern of charge transfer 
at the metal-molecule interface is similar for the two molecules, the 
long-range electrostatic potential perturbation is quite different. 
For the BPD molecule, the electrostatic potential profile inside the 
the two benzene rings becomes quite complicated which creates additional 
barrier for the electron motion within the molecule core. 
Due to the non-coplanar geometry, the barrier for injection from the 
right metal into the right half of the molecule is larger than 
that from the left metal to the left half (the right benzene ring 
gives a slightly smaller orbital overlap with the right electrode). 
Since the probability of electron tunneling through a potential barrier is 
sensitive to its shape, the charge-transfer induced potential change will 
have a profound influence on the electron transmission through the 
metal-molecule-metal junction. 
The presence of the potential barrier at the interface (and also in the 
molecule core for BPD) will affect the charge redistribution within the 
molecule when an additional field is applied, since it impedes 
the flow of electrons across the metal-molecule-metal junction. This will 
be discussed in the next section.

\subsection{Contact-induced modification of molecular states, band lineup 
and conductance of the metal-molecule-metal junction\label{S3-3}}

The coupling between the gold electrodes and the molecule (reflected in 
the self-energy operator of the electrodes) and the induced 
potential change in the the molecule (reflected in the self-consistent 
molecular Fock matrix) modify both the charge distribution and the 
energy of the molecular states relative to the metal Fermi-level. They also 
broaden the discrete molecular electronic spectrum into a quasi-continuous 
one. These effects can be analyzed through the local density 
of states, the transmission coefficient and their projection onto the 
individual molecular orbitals. 

The transmission versus energy (T-E) and the projected DOS (PDOS) 
corresponding to the five frontier molecular states of the two molecules 
(LUMO+1, LUMO, HOMO, HOMO-1 and HOMO-2) closest to the 
Fermi-level for spin-up electrons are shown in 
Figs. (\ref{xueFig3-3-1-1}) and (\ref{xueFig3-3-1-2}). 
The molecules studied here contain an even number of electrons and the 
gold electrode is non-magnetic, so the spin-up and spin-down channels show 
identical characteristics. For other molecules in contact with 
ferromagnetic electrodes, the spin-resolved analysis is essential for 
understanding the transport characteristics.  

From both the transmission versus energy curve and the projected DOS 
curve, we find the Fermi-level lies closer to the HOMO state than to the 
LUMO state for both molecules. Comparing the peak position 
in the PDOS plot with the energy levels of the isolated molecule, we find  
that the contact with the metallic electrodes significantly shifts the 
energy levels of the frontier molecular states. The change is found to 
be larger for the occupied states than for the unoccupied states since 
the unoccupied states are benzene based (which also leads to sharper 
peaks in the PDOS plot). 

The contact with the electrodes also modifies the charge distribution of the 
molecular states. This is illustrated by examining the LDOS at energies 
corresponding to the peak positions of the PDOS of the HOMO and LUMO 
for both molecules in Figs. (\ref{xueFig3-3-2-1}) and (\ref{xueFig3-3-2-2}). 
We show both the shape 
and the spatial distribution of the surface-perturbed molecular states 
for PDT. The spatial distribution is obtained by integrating the 
3-D LDOS with respect to the z-axis and plotted as a function of position 
in the xy-plane. The perturbation of the molecular states 
due to the coupling to the electrodes is different for the two 
molecules. In general, the surface-induced change in the charge 
distribution associated with the molecular states correlates closely 
with the change in the electrostatic potential. 
For the PDT molecule, the large increase of the electrostatic potential 
in the middle of the benzene rings pushes the electrons located on the 
carbon atoms to the peripheral hydrogen atoms for both the HOMO and the 
LUMO states (Fig. \ref{xueFig3-3-2-1}). The LUMO level remains highly 
localized giving rise to sharp peak in both PDOS and T-E plots. 
For the BPD molecule, electrons also move from the right benzene to the 
left benzene for both the HOMO and LUMO states due to the smaller 
potential arising from the difference in local geometries. 
Similar analysis applies to other molecular states.

The peak positions in the PDOS plots of the HOMO and LUMO states 
align closely with the two peak positions in the T-E plot 
around the Fermi-level, 
corresponding to the onset of resonant transmission. By projecting the total 
transmission coefficient onto the individual molecular orbitals, we find 
the first transmission peak above $E_{F}$ (at $E=-2.45(eV)$) arises mainly 
from the nearly 
degenerate LUMO and LUMO+1 states. The first transmission peak below $E_{F}$ 
(at $E=-6.5(eV)$) instead arises mainly from the HOMO state. 
In addition to the transmission peak at the HOMO and LUMO states, 
there is also a transmission 
peak in the HOMO-LUMO gap of the PDT molecule at energy $E=-3.6 (eV)$ 
(Fig. \ref{xueFig3-3-1-1}). This peak corresponds to transmission through 
metal-induced gap states (MIGS) due to the hybridization of metal 
surface states with the occupied molecular states, which is significant for 
such a short molecule as PDT. The projected transmission analysis shows 
contribution coming mainly from the HOMO and also other occupied molecular 
states. The nature of the MIGS state is most clearly seen in the 
corresponding LDOS plot (Fig. \ref{xueFig3-3-3}), which shows similar 
charge distribution to the HOMO state. This is the case for both the LDOS 
and the transmission coefficient throughout the HOMO-LUMO gap region 
including that at the Fermi-level. For such a short molecule as PDT, 
the transmission through the metal-induced gap states in the HOMO-LUMO 
gap is significant. A similar situatuation occurs also 
for the BPD molecule except that the transmission through the HOMO-LUMO gap 
of the BPD molecule is much reduced (Fig. \ref{xueFig3-3-1-2}). The molecule 
is longer and the orbital overlap with the right electrode states is 
weaker, so the transmission probability in the gap is much smaller. 

The magnitude of the transmission coefficient at the Fermi-level determines 
the zero-bias conductance at low temperature. We find a conductance 
of $4.8 (\mu S)$ and $1.4 (\mu S)$ for the PDT and BPD 
molecules. Since the length of the molecules are $6.4 (\AA)$ 
and $10.7 (\AA)$ respectively, the resistance is not proportional to   
the conjugation length, as expected for tunneling transport. 

\section{Device out of equilibrium\label{S4}}
 
\subsection{Current-voltage and conductance-voltage characteristics 
\label{S4-1}}

The current-voltage (I-V) and the differential conductance-voltage 
(dI/dV-V or G-V) characteristics are calculated for the two molecules 
using the method described in Sec.\ref{S2} in the bias range 
from $-4 (V)$ to $4 (V)$ and plotted in Figs. \ref{xueFig4-1-1} 
and \ref{xueFig4-1-2}. Due to the symmetric device structure, the 
current-voltage and conductance-voltage characteristics for both molecules 
are nearly symmetric with respect to bias polarity. In the low bias 
regime, current changes approximately linearly with the bias voltage for 
both molecules, corresponding to tunneling transport in the HOMO-LUMO gap. 
Both tunneling and thermionic emission contribution to the total current 
are shown. As expected, 
for tunneling transport through a large barrier, the thermionic 
emission contribution to the current is negligible even at low bias (shown 
in the insets). Since the variation of the thermionic emission 
contribution with bias voltage is small (in the coherent transport 
regime), this further reduces its 
contribution at high bias. Since this is the only temperature dependence 
in the coherent transport model, we expect the temperature dependence 
of the coherent current transport through the two molecules to be weak 
(ignoring any disorder effects). 
Within the coherent transport model, the thermionic emission and 
correspondingly the temperature dependence of the current transport can be 
important only if the the barrier for electron transmission is small.      

For comparison, we have also plotted the conductance-voltage 
characteristics obtained using the equilibrium transmission coefficient, 
.i.e., replacing $T(E,V)$ in Eq.\ \ref{IV} with $T(E,V=0)$. Since the gold  
surface density of states are approximately symmetric with respect to the 
Fermi-level (the gold surface band around the Fermi-level is due primarily 
to the $sp$ electrons), the difference between the G-V characteristics 
thus obtained and the self-consistent G-V characteristics reflects the 
effect of the bias-induced modification of molecular states. 

For both molecules, only the low-bias conductance-voltage characteristics 
(before reaching the first conductance peak) can be reasonably well 
reproduced by the equilibrium transmission characteristics. The deviation 
in both the magnitude and the peak position of the conductance becomes 
significant at large bias, there are also 
more conductance peaks than would be obtained from the equilibrium 
transmission characteristics. So any attempt to predicting the nonlinear 
transport characteristics from the equilibrium transmission characteristics 
combined with an assumption about the voltage drop will lead to significant 
error. The effect of the bias-induced modification of molecular states is 
also obvious from looking at the shift of the frontier molecular 
levels by the applied voltage, as shown in Fig.\ (\ref{xueFig4-1-3}). 
The molecular levels plotted are obtained by diagonalizing the molecular 
part of the self-consistent Fock matrices at each bias voltage. 
The molecular levels are nearly constant at low voltages, but begin to 
shift before reaching the first conductance peak. The voltage at which 
the shift occurs corresponds to the voltage where major deviation in 
the G-V characteristics occurs. 

An important question in current transport through molecules is which 
molecular states are responsible for the observed conductance peak. 
From the position of the plotted molecular levels, we expect that for both 
molecules, the peaks in the conductance are due to the 
\emph{occupied molecular} levels, in disagreement with previous calculation 
where a jellium model of the electrode is used~\cite{Lang002,Lang01}. 
We believe this is because the jellium model underestimates the metal work 
function which effectively pushes the HOMO level down relative to the 
metal Fermi-level~\cite{Xue01}. 
The shifts in the individual molecular levels are not identical in either 
direction or magnitude, so a rigid shift in energy levels as occuring in a 
planar metal-semiconductor contact does not apply here. Instead the 
shift of the individual molecular states will depend on the detailed 
charge and potential distributions. For the two molecules considered here, 
there is an effective reduction in the HOMO-LUMO gap as bias 
increases (Fig.\  \ref{xueFig4-1-3}).        
    
\subsection{Charge response and voltage drop 
\label{S4-2}}
The charge response of the molecule to the applied electrical field is 
reflected in both the net charge injection into the molecule and the 
charge redistribution inside the molecule. 
For PDT molecule, we find there is only slight additional charge 
injection into the 
molecule as we apply the bias voltage. For BPD molecules, the net 
charge injection becomes important only when we go to high bias ($>2.5(eV)$). 
The charge injection under nonzero bias voltage is determined by the balance 
of charge injection and extraction at the source-molecule and 
drain-molecule contacts. Since for the PDT molecule the contact geometries 
are identical and the gold surface density of states are approximately 
symmetric with respect to the Fermi-levels, little net charge accumulation 
will be induced by the applied bias voltages. 
For the BPD molecule, the two rings are non-coplanar, but due to 
the largely localized nature of the charge injection process, the difference 
in the two contact becomes non-negligible only at high bias voltages. 
In addition to the total net charge in the molecule, we can also calculate 
the nonequilibrium occupation of the individual orbitals from the 
nonequilibrium density matrix (Eq.\ \ref{DMNE}) as a functon of the bias 
voltage (Fig.\ \ref{xueFig4-2-1}). We find the change in the electron 
occupation is gradual at the voltages corresponding to the conductance 
peak, where molecular level moves past the fermi level of the electrodes. 
The electron occupation of the molecular orbital is fractional due to 
the broadening of the molecular level by the coupling to the contact.
  
Although the net charge injection due to applied voltage is negligible, 
there can be significant charge redistribution within the molecule.  
hiscalls for closer investigation into the spatial distribution of 
the charges and potential distribution within the molecule as a function 
of applied voltage~\cite{Mujica00}.   
Samples of the spatial distributions of the transferred charge and the 
electrostatic potential drop for both molecules at bias voltage of $3.0(V)$ 
are shown in Fig.\ \ref{xueFig4-2-2}. Similar 
charge and potential distributions are obtained for \emph{all} bias voltages.  
The spatial distribution of charge transfer is obtained by integrating 
the difference in electron density at finite and zero biases along the 
z-axis and plotted as a function of position in the xy-plane (defined 
by the left benzene ring). The potential drop is obtained by evaluating 
the difference between the electrostatic potential at finite and zero 
biases, which obeys the boundary condition of approaching $+V/2$ 
($-V/2$) at the left (right) electrode. The molecules exhibit 
different conductance at different voltages, but the calculated charge and 
potential distributions show similar patterns and \emph{do not depend 
on whether the molecules are in low or high conductance state}. The reason 
is as follows: The charge and potential response of the molecular junction 
is determined by the total electron population. But to be in high 
conductance state, one molecular orbital needs to align with one of the 
metal Fermi-levels. As it moves away from alignment, the change in the 
electron population is gradual (Fig.\ \ref{xueFig4-2-1}), in 
contrast with the change in conductance. Given the net 
charge injected into the molecule, the determination of the electrostatic 
charge and potential response of the molecule in contact with the 
two electrodes will be equivalent to that of an isolated \emph{charged} 
molecule under the same boundary conditions of charge and potential.  

The nature of the charge redistribution can be understood readily 
by partitioning the molecules into the atomic-groups, i.e., the benzene 
rings and the end sulfur atoms. As bias voltage increases, electrons move 
from the source-side carbon and sulfur to the drain-side carbon and sulfur. 
This is due to fact that the $\pi$ electrons in the carbon $Pz$ orbitals can 
move freely under the applied electric field. This flow of $\pi$ 
electrons is impeded at the molecule-drain contact because the the presence 
of potential barrier there, which inhibits the charge flow into the drain 
electrode. Similar considerations apply to the molecule-source contact. 
This leads to the creation of two resistivity dipoles at the metal-molecule 
interface, i.e., the accumulation of electrons on the injecting side and 
the depletion of electrons on the 
extracting side of the potential barrier. For the BPD molecule, where the two 
benzene benzene rings are non-coplanar, the charge response of the two 
benzene rings show identical behavior and are similar to the benzene 
ring in the PDT molecule. The weak orbital overlap between the two 
benzene rings disrupts the flow of $\pi$ electrons across the inter-benzene 
bonding region, which creates another potential barrier with 
corresponding resistivity dipole and partially insulates the two benzene 
rings from each other. 

The spatial variation of the current-induced electron redistribution could 
also have a significant effect on the structural stability of the molecule 
under high bias. As shown in Fig.\ \ref{xueFig4-2-2}, applying bias 
voltage increases the electron 
density in the sulfur-carbon bond at the source-molecule contact but 
decreases the electron density in the sulfur-carbon bond at the 
drain-molecule contact. According to Hellman-Feynman 
theorem~\cite{ForceBook}, this would lead to stronger attractive 
forces and consequently shorter sulfur-carbon bond at the source-molecule 
contact but weaker forces and longer sulfur-carbon bond at the 
drain-molecule contact. For the BPD molecule, there is also a shift of 
electron density in the carbon-carbon bond connecting the two benzene 
rings which may affect the inter-ring spacing. Since the magnitude of 
the electron redistribution increases with the bias voltage, we expect 
this bias-induced modification of structural change will have stronger 
effect at high-bias. The problem of current-induced conformational change 
is the molecular analogue of the more familiar electromigration problem in 
metallic systems~\cite{Verbruggen}. Although this problem has been 
treated recently by several groups~\cite{Lang01,Stafford99}, we feel 
that much needs to be done before a satisfactory theory emerges. 

The resistivity dipole is a well known 
concept in mesoscopic electron tranport~\cite{Landauer} and is common 
in inhomogenous transport media where their presence in the 
vicinity of local scattering center helps to overcome the barrier for 
transport and ensure current continuity. This can lead to a nonlinear 
transport effect due to the strong spatial variations in local carrier 
density and transport field. Such nonlinear transport characteristics induced 
by the device charge and potential inhomogeneity are a hallmark of band 
``engineering'' through mesoscopic semiconductor 
heterostructures~\cite{Wacker02}. Molecular junctions are an analogue of 
the mesoscopic ``heterostructures'', since the 
choice of the component functional and structural groups allows 
the possibility of ``engineering'' charge and potential inhomogeneity 
within a single molecule, suggesting the possibility of 
device ``engineering'' through molecular design and permitting the extension 
of device concepts to the molecular scale. This effect can be further 
elucidated by examining the potential response of the two molecules. 

The mobile $\pi$-electrons of the benzene ring effectively screen the 
applied field, leading to rather flat electrostatic potential drop across 
the benzene rings. The electrostatic potential drops rapidly in 
the molecule-electrode contact region, and becomes flat again as it 
approaches the boundary of the electrode where strong screening of the 
applied electric field occurs. For the PDT molecule, the 
majority of the voltage therefore drops across the molecule-electrode 
contact region~\cite{Datta01,Mujica00}. Since the unoccupied molecular 
states of the PDT molecule are located mainly on the benzene ring where the 
screened electric field is small, their energy 
levels don't change much as bias voltage increases. The 
HOMO-1 state shows stronger voltage dependence than the HOMO state 
because it is end-sulfur based where the potential 
variation is the strongest. For the BPD molecule, although the 
$\pi$ electrons of each benzene ring can screen the applied field 
effectively, the potential barrier between the two benzene rings 
prevents the electron from flowing freely across, leading to significant 
voltage drop across the carbon framework. As a result, there exists 
strong spatial variation of the electrostatic potential across the molecule 
core in addition to the metal-molecule contact. Since the four frontier 
molecular states of the BPD molecule are either sulfur-based or have 
charge distributed across the entire molecule core, they all show 
clear voltage dependence. An  empirical model of rigid shift of the 
molecular level relative to the Fermi-levels of the two electrodes has 
often been assumed~\cite{Datta97}, with the proportion of the voltage 
drop being associated with the contact geometry. 
It is clear that this model can only be utilized as a crude check of the 
device characteristics, since it neglects the possible voltage drop 
across the molecules and the different effects this may have on different 
molecular states. 

\subsection{Bias-induced modification of molecular states and 
transmission characteristics \label{S4-3}}

The molecular charge and potential response to the applied electrical field 
will affect the transport characteristics in two ways: (1) it shifts the 
molecular level relative to the Fermi-levels of the two electrodes; (2) it 
modifies the charge distribution of the molecular states and therefore 
their capability for carrying current. The bias-induced modification of 
molecular states can be analyzed by examining the local density of 
states and its projection onto the individual molecular orbitals at 
finite bias voltages. 

A snapshot of the bias dependence of the transmission characteristics 
and the projected DOS corresponding to the five frontier molecular 
orbitals is shown for the PDT molecule at bias voltages 
of $1.6(V),3.0(V),4.0 (V)$ 
(Fig.\ \ref{xueFig4-3-1}) and for the BPD molecule at bias voltages 
of $1.4(V),3.0(V),4.0(V)$ (Fig.\ \ref{xueFig4-3-3}) to illustrate their 
voltage dependence. 
We have also shown for PDT molecule the LDOS at energies corresponding 
to the peak position in the PDOS plots of the HOMO and LUMO states at bias 
voltage of $3.8(V)$ (Fig.\ \ref{xueFig4-3-2}). 
   
For the PDT molecule changing from zero bias (Fig.\ \ref{xueFig3-3-1-1}) 
to $V=1.6(V)$ (Fig.\ \ref{xueFig4-3-1}), both the peak positions in the 
PDOS and the electron distribution associated 
with the LUMO states don't change much. 
The HOMO energy level also doesn't change much, but there 
is a shift in charge distribution from the right end sulfur atom to the 
right carbon atom (not shown here). The energy of the HOMO-1 
state increases with bias, leading to a decrease in the energy spacing 
between the HOMO and HOMO-1 states. At $1.6(V)$, the transmission versus 
energy of the PDT molecules reaches a peak at energy corresponding to 
the drain Fermi-level at $E_{F}-eV/2$, giving rise to the first peak in 
the conductance. As we increase the bias voltage further from $V=1.6(V)$  
to $3.0(V)$ (Fig. \ref{xueFig4-3-1}), the energy levels of the HOMO 
and HOMO-1 states shift gradually toward the equilibrium Fermi-level 
$E_{F}$, increasing the transmission coefficient there. As we further increase 
the bias to $V=3.8(V)$, the charge distribution (the LDOS) at energies 
corresponding to the peak position of the LUMO state also changes, 
developing large weight on the end sulfur 
$P_{Z}$ and the peripheral hydrogen atoms (Fig. \ref{xueFig4-3-2}). 
Because the HOMO and HOMO-1 states move up to the metal Fermi-level 
as the bias increases, the current 
increases more gradually than would be obtained if the 
bias-induced modification of molecular states is neglected 
(Fig. \ref{xueFig4-1-1}). 
    
For the BPD molecule, the first peak in conductance is reached 
at $V=1.4(V)$ (Fig. \ref{xueFig4-1-2}) and Fig. \ref{xueFig4-3-3}). 
From zero bias to $1.4(V)$, the energies of the LUMO and HOMO-1 states 
decrease and there is a shift of charge from the right benzene to the 
left benzene ring. The energy of the HOMO states 
increases, and the shift of charge distribution is from the left benzene 
ring to the right benzene ring (not shown here). Compared to the 
PDT molecule, the bias-induced modification is stronger despite the 
longer molecule length, due to the stronger spatial 
variation of the potential profile. As we increase the bias to $V=3.0(V)$ 
(Fig. \ref{xueFig4-3-3}), the energy of the LUMO continues to 
decrease, but the energies of both the HOMO and HOMO-1 states increase 
toward the metal Fermi-level $E_{F}$. The 
transmission coefficient at both energies decreases significantly since 
the electrons are now more localized on the right ring. A notable 
change is that a second peak in the PDOS of the HOMO state develops 
at $E=-7.0 (eV)$. Examining the corresponding LDOS shows similar charge 
distribution to that of the HOMO-2 state with energy of $-7.15 (eV)$ 
which has charge distribution on both benzene rings (not shown here) and 
correspondingly large transmission probability. Increasing the bias 
to $4.0 (V)$ (Fig. \ref{xueFig4-3-3}) further decreases the energy of 
the LUMO state and increase the energies of the HOMO and HOMO-1 states. 
At $V=3.8(V)$, the electrons shift from the right benzne ring back to 
the left for both the HOMO and HOMO-1 states (not shown)
increasing the transmission coefficients there.   
Since the electrochemical potential of the left (right) electrode 
approaches alignment with the HOMO-2 and also the LUMO state, there is a 
rapid increase in the conductance as we increases the bias further toward 
$V=4.0 (V)$ (Fig. \ref{xueFig4-1-2}). 
 
\section{Discussion and conclusion \label{S5}}

\subsection{Electron transport or hole transport?}
A common practice in the literature on molecular electronics is to 
characterize the molecular transport as ``electron transport'' if the 
conduction is mediated by tunneling through the unoccupied molecuar states, 
or as ``hole transport'' if the conduction is mediated by tunneling through 
the occupied molecular states, following the terminology commonly used in 
semiconductor/organic device and electron transfer research. Here it is 
important to recognize their differences.

For bulk semiconductor devices, the concepts of ``electron'' or ``hole'' 
are associated with introducing shallow-impurity (dopant) atoms into the 
ideal lattice structure, which introduces electron intos the (unoccupied) 
conductand band or removes electrons from the (filled) valence band. 
The ``electron'' or ``hole'' thus introduced are mobile charge carriers 
with delocalized wavefunctions. The system as a whole remains charge neutral, 
since the charges associated with these carriers are compensated by the 
impurity ions left behind. Therefore, the type of charge transport is 
an equilibrium property depending only on the type and amount of dopant 
atoms introduced. By contrast, for organic devices, the system is often 
not doped. Charge transport occurs by injecting charge carriers 
into the system from the electrodes. The type of transport depends on the 
charging state of the molecule and the nonequilibrium injection of charge 
carriers into the neutral system. For both cases, the type of transport 
is determined by the change in the occupation of electron states through 
doping or contact injection, indpendent of the nature of charge transport 
itself (band transport or polaron hopping). 

The situation for coherent molecular transport is quite different. 
For the symmetric molecular devices considered here, both the total 
number of electrons (the charging state of the molecule) and the 
occupation of the individual molecular orbitals change little with 
the applied bias. Although a fractional amount of charge is transfered from 
the gold to the molecule upon electrode contact, the transfered charge is 
localized in the interfacial region and characterizes the changes in the 
interfacial bond. The change in the occupation of molecular states is 
fractional and gradual, either upon contact to the electrodes at equilibrium 
or upon application of a nonzero bias voltage out of 
equilibrium (Fig.\ (\ref{xueFig4-2-1})). This is due to the quantum 
mechanical nature of the coherent tunneling transport, 
where we cannot characterize the tunneling electron as being physically 
injected into the molecule and subsequently extracted out. Therefore, 
it is more appropriate to characterize molecular transport through the 
resonant molecular states without associating it with specific transport 
types.  

\subsection{Single molecule versus molecular monolayer}   

The focus of the present series of work is on current 
transport through a single molecule in contact with two metallic electrodes. 
This corresponds closely to the molecular transport measurements using 
atomic-size break junctions~\cite{Reed97,Weber02,Shash02,Mccuen02,Park02}, 
where either one or several molecules are probed. However, many molecular 
transport experiments are performed on molecular monolayers where 
thousands or tens of thousands of molecules are probed by the contacts to 
the electrodes~\cite{JGA00,Reed992,Tour01,Lindsay01}. Such experiments with 
monolayer configuration present a quite 
different situation from the single molecule configuration considered here. 
Besides the additional complexity of inter-molecular interactions within 
the monolayer, the most important difference lies in the interface 
electrostatics. Note that the same boundary condition for the electrostatic 
potential across the molecular junction applies to 
both the single molecule configuration and the monolayer configuration: 
deep inside the electrodes they approach the bulk value, which can be 
shifted rigidly with respect to each other by the applied bias voltage. 
For the single molecule configuration, the transfer of charge across the 
metal-molecule interface is confined in a small region, 
whose contribution to the electrostic potential decays to zero in regions 
far away from the molecule. But for the monolayer configuration, the 
electrostatic potential in regions far away from the molecule is the 
superposition of contributions from the transfered charge on a large 
number of molecules. To satisfy the same boundary condition of the  
electrostatic potential, the charge transfer per molecule in a molecular 
monolayer can be orders of magnitude smaller than that in the single 
molecule considered here~\cite{Note2,Note3}. 
This is the situation often met in organic electronics, where a 
monolayer of self-assembled molecules is used to modify the work function 
of the metallic contacts~\cite{Friend98,Marks99}. Similar problems have 
also been considered in the electron transport through metal-carbon 
nanotube interfaces~\cite{Tersoff99,Xue031}. Here it is important to 
recognize the different physics of the metal-molecule interface in 
single-molecule devices and monolayer devices, since it may have profound 
effects on the device functionalities achievable using molecular 
materials. 

\subsection{Conclusion}

We have presented a first-principles based microscopic study of current 
transport through individual molecules. The real-space formulation 
allows us to establish a clear connection between the transport 
characteristics and the molecular electronic structure perturbed by 
the metal-molecule coupling and the applied electric field. By 
separating the problem into equilibrium and non-equilibrium situations, 
we identify the critical electronic processes for understanding 
the linear and non-linear transport characteristics. At equilibrium, 
the critical problem is the charge transfer process 
upon formation of the metal-molecule-metal contact, which modifies the 
molecular states and determines the energy level lineup relative to the 
metal Fermi-level. This is mainly an interface-related process and 
can be controlled by controlling the contact. Out of 
equilibrium, the central problem is the molecular charge response and the 
consequent molecular screening of the applied electric field, 
which depends on both the molecule core and the nature 
of the metal-molecule contact and can be understood by viewing the 
molecules as ``heterostructures'' of chemically well-defined local groups 
and analyzing their electrical response to the applied electrical 
field~\cite{Note4}

\begin{acknowledgments}
We thank A. Nitzan, S. Datta, V. Mujica and H. Basch for useful discussions.
This work was supported by the DARPA Molectronics program, the DoD-DURINT 
program and the NSF Nanotechnology Initiative. 
\end{acknowledgments}



\begin{references}
\bibitem[*]{Xue} Author to whom correspondence should be addressed. 
Email: ayxue@chem.nwu.edu 
\bibitem{Ratner74} A. Aviram and M. A. Ratner, Chem.\ Phys.\ Lett.\ {\bf 29}, 
277 (1974). 
\bibitem{Ratner98} \emph{Molecular Electronics:  Science and Technology}, 
edited by A. Aviram and M. Ratner (New York Academy of Sciences, New York, 
1998).
\bibitem{JGA00} C. Joachim, J. K. Gimzewski and A. Aviram, Nature 
{\bf 408}, 541 (2000) and references thererein.
\bibitem{Reed97} M. A. Reed, C. Zhou, C. J. Muller, T. P. Burgin and 
J. M. Tour, Science {\bf 278}, 252 (1997); Mark A. Reed, 
Proc.\ IEEE {\bf 97}, 652 (1999) and references therein. 
\bibitem{Metzger97} R. M. Metzger et al.\ , 
J.\ Am.\ Chem.\ Soc.\ {\bf 119}, 10455 (1997); 
R. M. Metzger, Acc.\ Chem.\ Res.\ {\bf 32}, 950 (1999).
\bibitem{Reed992} J. Chen, M. A. Reed, A. M. Rawlett and J. M. Tour, 
Science {\bf 286}, 1550 (1999).  
\bibitem{Tour01} Z. J. Donhauser et al.\ , Science {\bf 292}, 2303 (2001).
\bibitem{DekkerR99} C. Dekker, Phys.\ Today, page 22, May 1999.
\bibitem{NTFET} R. Martel et al.\ , Appl.\ Phys.\ Lett.\ {\bf 73}, 2447 (1998);
H. T. Soh, \emph{ibid.} {\bf 75}, 627 (1999). 
\bibitem{Dekker00} D. Porath, A. Bezryadin, S. de Vries and C. Dekker, 
Nature {\bf 403}, 635 (2000). 
\bibitem{Datta97} S. Datta, W. Tian, S. Hong, R. Reifenberger, 
J. J. Henderson and C. P. Kubiak, Phys.\ Rev.\ Lett.\ {\bf 79}, 2530 (1997).
\bibitem{Xue991} Y. Xue, S. Datta, W. Tian, S. Hong, R. Reifenberger, 
J. J. Henderson and C. P. Kubiak, Phys.\ Rev.\ B.\ {\bf 59}, 7852 (1999).  
\bibitem{Hong00} S. Hong, R. Reifenberger, W. Tian, S. Datta, 
J. Henderson and C. P. Kubiak, Superlatt.\ Microstruc.\ {\bf 28}, 289 (2000).
\bibitem{Bao02} N. B. Zhitenov, H. Meng and Z. Bao, Phys.\ Rev.\ Lett.\ 
{\bf 88}, 226801 (2001).
\bibitem{Xue01} Y. Xue, S. Datta and M. A. Ratner, 
J.\ Chem.\ Phys.\ {\bf 115}, 4292 (2001).
\bibitem{Xue021} Y. Xue, S. Datta and M. A. Ratner, Chem.\ Phys.\ {\bf 281}, 
151 (2002). 
(LANL archive: xxx.lanl.gov/cond-mat/0112136). 
\bibitem{Datta01} P. S. Damle, A. W. Ghosh and S. Datta, Phys/\ Rev.\ B.\ 
{\bf 64}, 201403 (2001).
\bibitem{Lang002} M. Di Ventra, S. T. Pantelides and N. D. Lang, 
Phys.\ Rev.\ Lett.\ {\bf 84}, 979 (2000); N. D. Lang and Ph. Avouris, 
Phys.\ Rev.\ B {\bf 64}, 125323 (2001)
\bibitem{Lang01} M. Di Ventra and N. D. Lang, 
Phys.\ Rev.\ B {\bf 64}, 45402 (2001). 
\bibitem{Guo011} J. Taylor, H. Guo and J. Wang, 
Phys.\ Rev.\ B {\bf 63}, 245407 (2001). 
\bibitem{Brandbyge02} M. Brandbyge et al.\ , Phys.\ Rev.\ B 
{\bf 65}, 165401 (2002).
\bibitem{Nitzan} A. Nitzan, Annu.\ Rev.\ Phys.\ Chem.\ {\bf 52}, 681 (2001).
\bibitem{EK99} E. G. Emberly and G. Kirczenow, Phys.\ Rev.\ B {\bf 60}, 6028.
\bibitem{Hush00} L.\ E.\ Hall et al.\ , J.\ Chem.\ Phys.\ {\bf 112}, 
1510 (2000). 
\bibitem{Semi} J. M. Seminario, A. G. Zacarias and J. M. Tour, 
J.\ Am.\ Chem.\ Soc.\ {\bf 121}, 411 (1999).  
\bibitem{Ho98} B. C. Stipe, M. A. Rezaei and W. Ho, Science {\bf 280}, 1732 
(1998); Phys.\ Rev.\ Lett.\ {\bf 82}, 1724 (1999). 
\bibitem{Yu99} Z. G. Yu, D. L. Smith, A. Saxena and A. R. Bishop, 
Phys.\ Rev.\ B {\bf 59}, 16001 (1999).
\bibitem{Fisher01} H. Ness, S. A. Shevlin and A. J. Fisher, Phys.\ Rev.\ B 
{\bf 63}, 125422 (2001). 
\bibitem{Keldysh65} L. V. Keldysh, Sov. Phys. JETP {\bf 20}, 1018 (1965).
\bibitem{Langreth76} D. C. Langreth, 
in \emph{Linear and Non-linear Electron Transport in Solids}, 
edited by J. T. Devreese and E. Van Doren (Plenum Press, New York, 1976). 
\bibitem{Daniel84} P. Danielewicz, Ann. Phys. {\bf 152}, 239 (1984). 
\bibitem{Mahan87} G. D. Mahan, Phys.\ Rep.\ {\bf 145}, 251 (1987).
\bibitem{HJBook} H. Haug and A-P. Jauho, \emph{Quantum Kinetics in 
Transport and Optics of Semiconductors} (Springer-Verlag, Berlin, 1996). 
\bibitem{Wilk99} W. G. Aulbur, L. J\"{o}nsson and J. W. Wilkins, 
Solid State Phys.\ {\bf 54}, 1 (1999).
\bibitem{LMBook} \emph{Theory of The Inhomogeneous Electron Gas}, 
edited by S. Lundqvist and N. H. March (Plenum Press, New York, 1983). 
\bibitem{DGBook} R. M. Dreizler and E. K. U. Gross, \emph{Density 
Functional Theory: An Approach to the Quantum Many-Body Problem} 
(Springer-Verlag, Berlin, 1990). 
\bibitem{Hoffmann88} R. Hoffmann, Rev.\ Mod.\ Phys.\ {\bf 60}, 601 (1988). 
\bibitem{ABW85} T. A. Albright, J. K. Burdett and M.-H. Whangbo, 
\emph{Orbital Interactions in Chemistry} 
(John Wiley and Sons, New York, 1985).
\bibitem{Lang82} A. R. Williams, P. J. Feibelman and N. D. Lang, 
Phys. Rev. B {\bf 26}, 5433 (1982). 
\bibitem{BT99} G. P. Brivio and M. I. Trioni, Rev.\ Mod.\ Phys.\ 
{\bf 71}, 231 (1999).
\bibitem{Ratner99} S. N. Yaliraki, A. E. Roitberg, C. Gonzalez 
and M. A. Ratner, J.\ Chem.\ Phys.\ {\bf 111}, 6997 (1999). 
\bibitem{Joachim97} M. Magoga and C. Joachim, 
Phys.\ Rev.\ B {\bf 56}, 4722 (1997). 
\bibitem{Dede82} R. Zeller, J. Deuta and P. H. Dederichs,  
Solid State Commun.\ {\bf 44}, 993 (1982). 
\bibitem{Becke88CP} A. D. Becke, J.\ Chem.\ Phys.\ {\bf 88}, 2547 (1988).
\bibitem{MW92} Y. Meir and N. S. Wingreen, 
Phys.\ Rev.\ Lett.\ {\bf 68}, 2512 (1992). 
\bibitem{Datta95} S. Datta, \emph{Electron Transport in Mesoscopic systems}, 
(Cambridge University Press, Cambridge, 1995)
\bibitem{Cuevas98} J. C. Cuevas, A. Levy Yeyati, and A. Martin-Rodero, 
Phys.\ Rev.\ Lett.\ {\bf 80}, 1066 (1998). 
\bibitem{Becke88GGA} A. D. Becke, Phys.\ Rev.\ A.\ {\bf 38}, 3098 (1988).
\bibitem{PW91} J. P. Perdew and Y. Wang, Phys.\ Rev.\ B.\ {\bf 38}, 
12228 (1988); M. Ernzerhof, J. P. Perdew and K. Burke, in \emph{Density 
Functional Theory I}, edited by R. F. Nalewajski (Springer, Berlin, 1996). 
\bibitem{JG89} O. Gunnarson and B. I. Lundqvist, Phys.\ Rev.\ \ B 
{\bf 13}, 4274 (1976); R. O. Jones and O. Gunnarson, Rev.\ Mod.\ Phys.\ 
{\bf 61}, 689 (1989). 
\bibitem{KBT} L. R. Kahn, P. Bybutt and D. G. Truhlar, 
J.\ Chem.\ Phys.\ {\bf 65}, 3826 (1976).
\bibitem{Stevens84} For carbon and sulfur, we use the pseudopotential and the 
corresponding polarized split valence basis sets of W. J. Stevens, H. Basch 
and M. Krauss, J.\ Chem.\ Phys.\ {\bf 81}, 6026 (1984); For gold, we use the 
pseudopotential and the valence basis sets of P. J. Hay and 
W. R. Wadt, J.\ Chem.\ Phys.\ {\bf 82}, 270 (1985).
\bibitem{Note1} For the gold atoms lying on top of the semi-infinite 
substrate (e.g. describing the atomic-scale structure of the surface), we 
use the standard double-zeta basis set as described by P. J. Hay and 
W. R. Wadt (J.\ Chem.\ Phys.\ {\bf 82}, 270 (1985)); For the gold atoms 
on the surface of the semi-infinite substrate, we follow the standard 
convention in surface and solid state electronic structure calculation 
using molecular Gaussian-type orbital basis sets to construct a 
minimum valence basis set by removing the most diffuse Gaussian primitive. 
A minimum-valence basis set 
may also be formed including the most diffuse component 
(H. Basch, unpublished), however we find 
this leads to convergence problem and overcharging of the ``extended 
molecule'' with charge overflow onto the substrate atoms without affecting 
the molecule significantly. For further information 
rearding the basis set usage for surface and bulk calculations, see 
the seris of work related to the CRYSTAL98 package (V. Saunders, R. Dovesi, 
C. Roetti, M. Causa, N. Harrison, R. Orlando, and C. M. Zicovich-Wilson, 
Universit{\`{a}} di Torino, Torino, 1998), e.\ g.\ , Y. Noel et al.\ , 
Phys.\ Rev.\ B {\bf 65}, 14111 (2001); A. Kokaly and M. Causa, 
J.\ Phys.: Condens.\ Matter {\bf 11}, 7463 (1999). 
\bibitem{Papa86} D. A. Papaconstantopoulos, \emph{Handbook of the Band 
Structure of Elemental Solids} (Plenum Press, New York, 1986).
\bibitem{G98} GAUSSIAN 98, Revision A.7, 
M. J. Frisch, G. W. Trucks, H. B. Schlegel,  G. E. Scuseria, 
M. A. Robb, J. R. Cheeseman, V. G. Zakrzewski, J. A. Montgomery Jr.\ ,  
 R. E. Stratmann, J. C. Burant,  S. Dapprich, J. M. Millam,  
 A. D. Daniels, K. N. Kudin,  M. C. Strain, O. Farkas, J. Tomasi,  
 V. Barone, M. Cossi, R. Cammi, B. Mennucci, C. Pomelli, C. Adamo,  
 S. Clifford, J. Ochterski,  G. A. Petersson, P. Y. Ayala, Q. Cui,  
 K. Morokuma, D. K. Malick, A. D. Rabuck, K. Raghavachari,  
 J. B. Foresman, J. Cioslowski, J. V. Ortiz, A. G. Baboul,  
 B. B. Stefanov, G. Liu, A. Liashenko, P. Piskorz,  
I. Komaromi, R. Gomperts, R. L. Martin, D. J. Fox, T. Keith, M. A. Al-Laham,  
 C. Y. Peng, A. Nanayakkara, C. Gonzalez, M. Challacombe,  
 P. M. W. Gill, B. Johnson, W. Chen, M. W. Wong, J. L. Andres,  
 C. Gonzalez, M. Head-Gordon, E. S. Replogle and J. A. Pople", 
 Gaussian, Inc.\ , Pittsburgh, PA, 1998. 
\bibitem{Heine65} V. Heine, Phys.\ Rev.\ A{\bf 138}, 1689 (1965).
\bibitem{Tersoff} J. Tersoff, Phys.\ Rev.\ Lett.\ {\bf 52}, 465 (1984);
J. Tersoff, Phys.\ Rev.\ B {\bf 30}, 4874 (1984).
\bibitem{Monch90} W. M{\"{o}}nch, Rep.\ Prog.\ Phys.\ {\bf 53}, 221 (1990).
\bibitem{MSBook} E. H. Rhoderick and R. H. Williams,  
\emph{Metal-Semiconductor contact}, 2nd edition 
(Oxford University Press, Oxford, 1988).
\bibitem{CS01} I. H. Campbell and D. L. Smith, Solid State Phys.\ 
{\bf 55},1 (2001).
\bibitem{Friend98} P. K. H. Ho, M. Granstr{\"{o}}m, R. H. Friend and 
N. C. Greenham, Adv.\ Mat.\ {\bf 10}, 769 (1998).
\bibitem{Marks99} W. Li et al.\ , Adv.\ Mat.\ {\bf 11}, 730 (1999).
\bibitem{Mujica00} V. Mujica, A. E. Roitberg and M. A. Ratner, 
J.\ Chem.\ Phys.\ {\bf 112}, 6834 (2000). 
\bibitem{Landauer} R. Landauer, IBM J.\ Res.\ Develop.\ {\bf 1}, 223 (1957); 
{\bf 32}, 306 (1988).  
\bibitem{Wacker02} A. Wacker, Phys.\ Rep.\ {\bf 357}, 1 (2002). 
\bibitem{ForceBook} \emph{"The Force Concept in Chemistry}, edited by 
B. M. Deb (Van Norstrand Reinhold Company, New York, 1981).
\bibitem{Verbruggen} A. H. Verbruggen, IBM J.\ Res.\ Develop.\ 
{\bf 32}, 93 (1988). 
\bibitem{Stafford99}  F. Kassubek, C. A. Stafford and H. Grabert, 
Phys.\ Rev.\ B {\bf 59}, 7560 (1999).
\bibitem{Weber02} J. Reichert et al.\ , Phys.\ Rev.\ Lett.\ {bf 88}, 
176804 (2002).  
\bibitem{Shash02} J. G. Kushmerick et al.\ , Phys.\ Rev.\ Lett.\ {bf 89}, 
86802 (2002).  
\bibitem{Mccuen02} J. Park et al.\ , Nature {\bf 417}, 722 (2002).
\bibitem{Park02} W. Liang et al.\ , Nature {\bf 417}, 725 (2002). 
\bibitem{Lindsay01} X.D. Cui et al.\ , Science {\bf 294}, 571 (2001). 
\bibitem{Note2} A simple estimate can be given in the limit of an ideal 
(infinite)  two-dimensional molecular monolayer. We assume the 
molecule-metal distance is $3(\AA)$ and each molecule occupies a 
surface area of $10(\AA ^{2})$ on average. 
Replacing the charge distribution of the two-dimensional monolayer 
by an uniform charge sheet, a charge transfer of $0.02$ per molecule 
is enough to give an electrostatic potential drop of $2.5(V)$ across 
the metal-monolayer junction.
\bibitem{Note3} D. Janes, private communication. 
\bibitem{Tersoff99} F. L{\'{e}}onard and J. Tersoff, Phys.\ Rev.\ Lett.\ 
{\bf 83}, 5174 (1999).
\bibitem{Xue031} Y. Xue and M.A. Ratner, to be published. 
\bibitem{Note4} Since our emphasis in this work is in conceptual issues, 
we have not tried testing or optimizing the more technical 
aspects of the modeling work such as the choice of the exchange-correlation 
functionals and the optimum molecular basis sets. For example, the 
calculations reported in this work were performed using a quite 
restricted basis set for the metal atoms on the 
surface of the semi-infinite substrate. The metal atoms belonging to 
atomic-scale structures on top of the substrate and the molecule itself are 
treated using much better basis sets as in standard molecular calculations. 
It is clear that the description of the metallic substrate is inadequate. 
And obtaining a better balance in the basis set decription between the 
substrate atoms and the molecule is a difficult and important issue for 
further improving the accuracy of the modeling of molecular-scale devices 
within the NEGF formalism. Such refinement may be needed for obtaining 
quantitative agreement with specific transport measurement. In particular, 
the calculated charger transfer from the gold surface to sulfur may 
be different depending on the basis sets (plane waves or extended gaussian 
basis sets) as well as the surface models (semi-infinite crystal, cluster 
or supercell geometry) chosen.              
 
\end{references}


%



\begin{figure}

\caption{\label{xueFig2-1-1} 
Atomic geometry of the gold-PDT-gold junction. 
Six gold atoms closest to the end sulfur 
atoms on each electrode are included into the ''extended molecule''. }
\end{figure} 
  
\begin{figure}
\caption{\label{xueFig2-1-2} 
Atomic geometry of the gold-BPD-gold junction. 
Six gold atoms closest to the end sulfur 
atoms on each electrode are included into the ''extended molecule''. }
\end{figure} 

 \begin{figure}
\caption{\label{xueFig3-1-1} 
Orbital shape of the HOMO-1, HOMO, LUMO and LUMO+1 states of PDT}
\end{figure}   

\begin{figure}
\caption{\label{xueFig3-1-2}
Orbital shape of the HOMO-1, HOMO, LUMO and LUMO+1 states of BPD}
\end{figure}   

\begin{figure}
\caption{\label{xueFig3-2-1}
Charge transfer upon the formation of the gold-PDT-gold contact. 
The top and the middle figure show the isosurface plot of the region where 
electron increases and decreases respectively. The bottom figure shows the 
spatial distribution of the transfered electrons as a function of position 
in the xy-plane after integrating over the z-axis. }
\end{figure} 
  
\begin{figure}
\caption{\label{xueFig3-2-2}
Charge transfer upon the formation of the gold-BPD-gold contact. 
The top and the middle figure show the isosurface plot of the region where 
electron increases and decreases respectively. The bottom figure shows the 
spatial distribution of the transfered electrons as a function of position 
in the xy-plane after integrating over the z-axis. }

\end{figure}   

\begin{figure}
\caption{\label{xueFig3-2-3-1}
Electrostatic potential change upon the formation of the 
gold-PDT-gold contact as a function of position in the xy-plane. Also shown 
is the projection of the molecule onto the xy-plane.}
\end{figure} 

\begin{figure}
\caption{\label{xueFig3-2-3-2}
Electrostatic potential change upon the formation of the 
gold-BPD-gold contact as a function of position in the xy-plane. Also shown 
is the projection of the molecule onto the xy-plane.}
\end{figure}   

\begin{figure}
\caption{\label{xueFig3-3-1-1} 
Band lineup at the gold-PDT-gold contact. The curves corresponding to the 
spin-up and spin-down electrons are virtually identical. 
The left figure plots the transmission versus energy, while the right 
figure plots the projected density of states onto the five frontier 
molecular states which can be identified from their peak positions. 
For PDT, these are -2.25(eV) (LUMO+1), -2.45(eV) (LUMO), 
-6.5(eV) (HOMO), -7.25(eV) (HOMO-1) and -7.7(eV) (HOMO-2). The horizontal 
line corresponds to the energy levels of four frontier molecular orbitals 
as plotted in Fig.\ \ref{xueFig3-1-1}. 
Note that the sharp peaks in the PDOS plot have been truncated here because 
showing them in full would have made the peaks corresponding to HOMO and 
LUMO less visible. }
\end{figure}

\begin{figure}
\caption{\label{xueFig3-3-1-2} 
Band lineup at the gold-BPD-gold contact. The curves corresponding to the 
spin-up and spin-down electrons are virtually identical. 
The left figure plots the transmission versus energy, while the right 
figure plots the projected density of states onto the five frontier 
molecular states which can be identified from their peak positions. 
For BPD, these are -2.5(eV) (LUMO+1), -2.85(eV) (LUMO), -6.35(eV) (HOMO),
-6.95(eV) (HOMO-1) and -7.45(eV) (HOMO-2). The horizontal 
line corresponds to the energy levels of four frontier molecular orbitals 
as plotted in Fig.\ \ref{xueFig3-1-2}. }
\end{figure}   

\begin{figure}
\caption{\label{xueFig3-3-2-1}
Characteristics of the surface perturbed HOMO and LUMO 
molecular states at the gold-PDT-gold contact. }
\end{figure}   

\begin{figure}
\caption{\label{xueFig3-3-2-2}
Characteristics of the surface perturbed HOMO and LUMO 
molecular states at the gold-BPD-gold contact. }
\end{figure}   
   

\begin{figure}
\caption{\label{xueFig3-3-3}Characteristics of the metal induced gap states. 
Left figure: gold-PDT-gold contact. Right figure: gold-BPD-gold contact.}
\end{figure}  
 
\begin{figure}
\caption{\label{xueFig4-1-1} 
I-V (upper figure)and G-V (lower figure) characteristics of the 
gold-PDT-gold device. The inset in the I-V plot gives the 
maginified view at low bias. The dotted line in the G-V plot is obtained 
assuming the transmission-energy relation to be bias-independent. }
\end{figure}   

\begin{figure}
\caption{\label{xueFig4-1-2}
I-V (upper figure) and G-V (lower figure) characteristics of the 
gold-BPD-gold device as in Fig.\ \ref{xueFig4-1-1}.}
\end{figure}
   
\begin{figure}
\caption{\label{xueFig4-1-3} 
Bias-induced modification of molecular levels at gold-PDT-gold junction 
(left figure) and gold-BPD-gold junction (right figure). We have also 
shown the position of the equlibrium Fermi-level $E_{F}$ and the 
electrochemical potential of the two electrode $\mu _{L(R)}$ in the plot.}
\end{figure}   

\begin{figure}
\caption{\label{xueFig4-2-1}
Nonequilibrium occupation of molecular orbitals as a function of voltage 
for the PDT (left figure) and BPD (right figure) 
molecules in the molecular junction. Here we show the 
electron occupation of the HOMO-1, HOMO, LUMO and LUMO+1 states. }
\end{figure}   
   
\begin{figure}
\caption{\label{xueFig4-2-2}
Spatial distribution of charge transfer and potential drop 
at the gold-PDT-gold and gold-BPD-gold contacts for bias voltage of $3.0(V)$. }
\end{figure}   

\begin{figure}
\caption{\label{xueFig4-3-1}Bias-induced modification of molecular 
states and transmission coefficent at voltages of 
$1.6(V),3.0(V)$ and $3.8(V)$ for the gold-PDT-gold contact. The sharp peaks 
in the PDOS plot are not shown in full here. }
\end{figure}    

\begin{figure}
\caption{\label{xueFig4-3-2}Characteristics of field-induced modification 
of molecular states at $V=3.8(V)$ for the gold-PDT-gold junction. 
Left figure (right figure) shows the LDOS at energy corresponding to the peak 
position in the projected DOS of the LUMO (HOMO) states. }
\end{figure}   
 
\begin{figure}
\caption{\label{xueFig4-3-3}
Bias-induced modification of molecular states and transmission 
coefficent at voltages of $1.4(V),3.0(V)$ and $4.0(V)$ 
for the gold-BPD-gold contact. The sharp peaks 
in the PDOS plot are not shown in full here.}
\end{figure}   

\end{document}